\documentclass[aps, prb, twocolumn, showpacs, floatfix,10pt,superscriptaddress]{revtex4-1}
\usepackage{amsmath}
\usepackage{amsfonts}
\usepackage{amsbsy}
\usepackage{amssymb}
\usepackage{graphicx}
\usepackage{textcomp}
\usepackage[caption=false,singlelinecheck=false]{subfig}
\usepackage{xcolor}
\usepackage{mathrsfs}
\usepackage{mathtools}
\usepackage{bm}
\usepackage{gensymb}
\usepackage{braket,wasysym}
\usepackage[colorlinks,linkcolor=blue,citecolor=red,filecolor=magenta,urlcolor=cyan,breaklinks]{hyperref}
\usepackage[bottom]{footmisc}
\usepackage{verbatim}
\DeclareMathOperator{\sech}{sech}
\hypersetup{colorlinks=true, urlcolor=blue, citecolor=cyan, pdfborder={0 0 0}}
\usepackage{breakurl}
\usepackage{natbib}

\allowdisplaybreaks

\begin{document}
\title{Phonon transmittance of one dimensional quasicrystals}

\author{Junmo Jeon}
\email{junmo1996@kaist.ac.kr}
\author{SungBin Lee}
\email{sungbin@kaist.ac.kr}

\affiliation{Korea Advanced Institute of Science and  Technology, Daejeon, South Korea}

\date{\today}
\begin{abstract}
In quasicrystals, special tiling patterns could give rise to unique physical phenomena such as critical states distinct from periodic systems. In this paper, we study how quasi-periodicity in aperiodic systems results in anomalous phonon modes, especially focusing on thermal transmittance in one-dimensional quasicrystals. Unlike periodic or compeletly random systems, we classify certain quasicrystals could host critical phonon modes whose transport properties are topologically protected based on their pattern equivariant cohomology group of supertilings. Starting from discussing general rule to find such critical phonon modes, we discuss classification of topologically distinct thermal transmittance in quasiperiodic systems. To be more specific, we exemplify  (\textit{decorated}) metallic-mean tilings and Cantor tiling, and derive universal features for resonant and decaying phonon modes as a function of quasi-periodic strength. Our study paves a new way to understand thermal transmittance of quasi-periodic systems based on the topological classification and offers quasicrystals as strong candidates to control drastic phonon modes.
\end{abstract}
\maketitle

\textit{\textbf{Introduction ---}} 
A phonon mode in condensed matter physics has been studied for a long time in the context of both thermal and electrical properties of materials
\cite{Simon:1581455,kittel1976introduction,cohen2016fundamentals,chaikin1995principles,bilz2012phonon}.
In particular, for a periodic system, presence of unit length scale makes thermal transmittance to be simply understood by phonon modes based on the Bloch's theorem\cite{kittel1976introduction,Simon:1581455,cohen2016fundamentals}. Whereas, in the aperiodic system, absence of periodic length scale makes such analysis impossible and thus study of thermal properties is quite challenging in general. 
Instead, when the system forms a quasi-periodic pattern, one could expect it's quasi-periodicity may give rise to unique phonon modes and  lead to a new type of thermal transmittance\cite{suck2013quasicrystals,steinhardt1987physics,janot2012quasicrystals,macia2000thermal}.

Historically, there are various discoveries in thermal properties of quasicrystals\cite{steinhardt1987physics,kawazoe2003structure,suck2013quasicrystals,macia2000thermal,bruin2013renormalization}. Both analytic and numerical studies have been performed related to thermal conductivity\cite{archambault1997thermal,steinhardt1987physics,gianno2000low,macia2000thermal,socolar1986phonons,luck1986phonon,zhu1999phonon,kohmoto1986quasiperiodic}.
In Fibonacci quasicrystal, for instance, the fractal structure of phonon density of states and existence of bizarre thermal transmittance characterized by neither extended nor localized phonon modes have been investigated using transfer matrix renormalization technique\cite{krajvci1995phonon,macia2000thermal,silva2019phononic}.
However, how can one classify distinct thermal transport properties for many quasi-periodic systems? 
In addition, understanding the controllability of thermal transmittance in quasicrystals remains  another important question we should ask.

In this paper, we explore the critical phonon modes in one dimensional quasicrystals and classify them for distinct topological properties. Based on the transfer matrix renormalization technique and the pattern equivariant (PE) cohomology of supertiling, our study develops a general scheme to classify unique thermal transmittance in quasicrystals. To illustrate our idea, we exemplify metallic-mean quasicrystals and the Cantor tiling and first derive the condition for presence and absence of critical phonon modes at special  frequencies. It turns out that the system does not possess any critical phonon mode for the original metallic-mean quasicrystal case. However, for some of the \textit{decorated} metallic-mean quasicrystals,
we find unique critical phonon modes exist. In this case, the corresponding thermal transmittance shows anomalous behavior in a sense that two different modes either critical or localized are drastically controlled by quasi-periodic strength. 
We emphasize that such anomalous thermal transmittance is not a general feature for arbitrary aperiodic tilings, but entirely depends on distinct PE cohomology classification of supertilings. 
We support this argument by showing other examples including the Cantor tiling case that belongs to different PE cohomology group of supertilings  and only exhibits rapidly decaying thermal transmittance. Our study guides the new plethora to understand anomalous thermal transmittance of quasicrystals that is absolutely absent in a periodic system, and to classify them based on the PE cohomology group. 

\textit{\textbf{Supertilings for phonon modes ---}} \ 
\textit{Supertiling} technique is the method of reconstructing patterns for local patches, which is generally applicable for studying electrical or thermal properties\cite{jeon2020topological}. 
To be more specific, let's consider the simple model of one-dimensional phonons assuming that atoms are connected with the nearest neighbor atoms. Then, the motion of each site is connected by two interaction links, left and right sides and the equation of motion is described by local transfer matrix $M_{[n-1, n+1]}$ as below\cite{kohmoto1986quasiperiodic,steinhardt1987physics},
\begin{align}
\label{eq:transfereq}
&\begin{pmatrix} u_{n+1}  \\ u_{n} \end{pmatrix}=\begin{pmatrix} \frac{K_{n}+K_{n-1}-m_n\omega^2}{K_n} & -\frac{K_{n-1}}{K_n} \\ 1 & 0 \end{pmatrix}\begin{pmatrix} u_n \\ u_{n-1} \end{pmatrix}.
\end{align}
Here, $K_n$ is the spring constant between site $n$ and site $n+1$. $u_n$ and $m_n$ is displacement of lattice vibration and mass on site $n$ respectively and $\omega$ is a mode frequency.

\begin{figure}[]
\centering
\includegraphics[width=0.5 \textwidth]{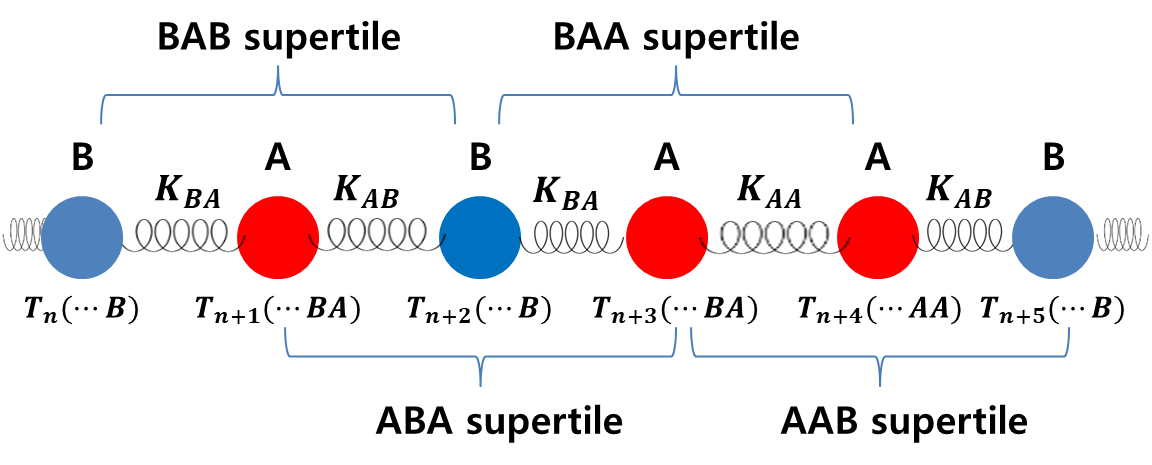}
\caption{\label{fig:transmittanceassign} 
Schematic picture to show definitions of spring constants, four kinds of length 2 supertiles $BAB,ABA,BAA,AAB$ in the \textit{induced} Fibonacci quasicrystal. There are three types of thermal transmittance $T_n (\cdots B),T_n(\cdots BA)$ and $T_n(\cdots AA)$ which distinguish sites $n$ and $n-1$ are occupied by either $A$ or $B$.}
\label{fig:supertiles}
\end{figure}

Now, one can group them with length 2 supertiles. (For instance, Fig.\ref{fig:supertiles} illustrates $BAB$, $BAA$, $ABA$, $AAB$ supertiles.) 
With setting of supertiles, one can further renormalize supertilings into multiple repetitive local patches. Particularly for certain phonon modes where the transfer matrices of such renormalized supertiles commute with each other, transmittance shows  topologically protected behavior which will be discussed later and we dub such a special phonon mode as a critical phonon mode\cite{macia2000thermal,steinhardt1987physics}.
For non-commuting case, however, 
transmittance behavior is generally not a topological quantity but sensitively depends on microscopic details. 
Hence, focusing on critical phonon modes where the transfer matrices of renormalized supertiles perfectly commute, we discuss universal features of thermal transmittance.

\textit{\textbf{Absence of critical phonon modes ---}} \ Following the above discussion, we first exemplify  the original Fibonacci quasicrystal and show the critical phonon modes are absent i.e., the transfer matrices for renormalized supertiles do not generally commute. (Later, we will discuss the cases for other quasi-periodic systems as well.) 
Let's first consider that quasi-periodicity lies on links only in terms of two prototiles $L$ and $S$. The substitution rule $L\!\to\!LS$, $S\!\to\!L$ leads to only three types of length 2 supertiles; $LS$, $SL$ and $LL$ (no $SS$) and the transfer matrices for each case are defined as  $M_{LS}$, $M_{SL}$ and $M_{LL}$ respectively. (See Supplementary Material for details.) 
Then, for the first few tiles starting with $LSLL\cdots\ $, the phonon motion is represented by the consecutive products of the transfer matrices $\cdots M_{LL}M_{SL}M_{LS}$. 
Based on the tiling pattern, there are certain rules; $M_{LL}$ is followed by $M_{SL}$ while $M_{SL}$ is followed by $M_{LS}$. Thus, one can express transmittance in terms of further renormalized transfer matrices, $M_{LL}$ and $M_{SL}M_{LS}$ which satisfy the following commutation relation, 
\begin{align}
\label{eq:commutator}
&[M_{LL},M_{SL}M_{LS}]=\lambda(\gamma^{-1}-1)\sigma_x. 
\end{align}
Here, $\lambda=\frac{m\omega^2}{K_L}$, $\gamma=\frac{K_S}{K_L}$ and $\sigma_x$ is the Pauli matrix.
From Eq.\eqref{eq:commutator}, it is clear that $M_{LL}$ and $M_{SL}M_{LS}$ do not commute with each other and thus a nontrivial critical mode is absent. (Note that $\gamma =1$ and $\lambda=0$ are the trivial limits for periodic pattern and $\omega=0$ respectively.) 
In similar manner, one can prove the absence of critical modes for other types of quasi-periodic cases which is discussed in the supplementary material.

\begin{figure}[!h]
\centering
\subfloat[]{\includegraphics[width=0.45 \textwidth]{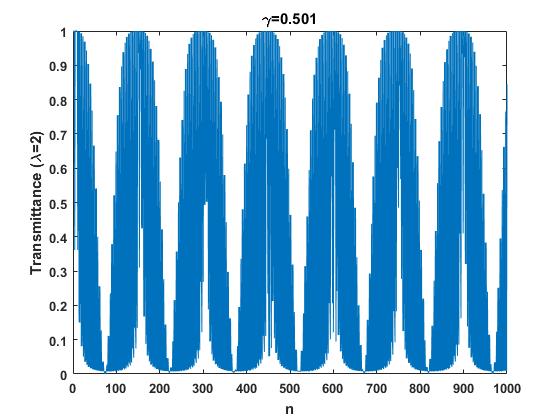}
\label{fig:501inducedFIB}}
\hfill
\subfloat[]{\includegraphics[width=0.45 \textwidth]{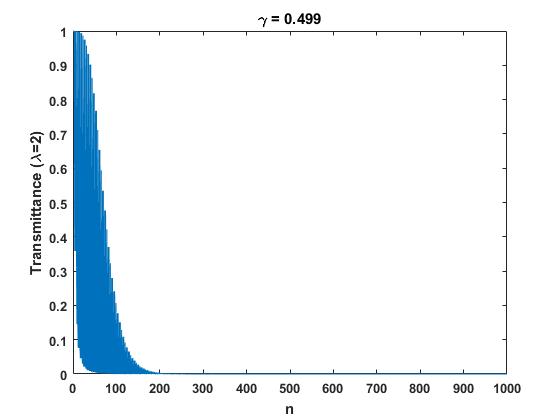}
\label{fig:499inducedFIB}}
\caption{\label{fig:inducedFIB} 
Thermal transmittance of the \textit{induced} Fibonacci quasicrystals for $\gamma = 0.501$ (a) and $\gamma=0.499$ (b) at $\lambda=2$ and $\alpha=1$. For $\gamma=0.501$, the critical phonon mode described by neither perfectly delocalized nor localized appears, whereas for $\gamma=0.499$ the localized phonon mode appears. See main text for details. }
\end{figure}

\textit{\textbf{Presence of critical phonon modes ---}} \ 
Unlike the original Fibonacci quasicrystal, decoration of quasi-periodicity in the Fibonacci tiling allows the presence of critical modes. 
Explicitly, we exemplify the case where quasi-periodicity exists both on sites and links simultaneously, termed as an \textit{induced} Fibonacci quasicrystal. In this case, sites are arranged as Fibonacci pattern, $ABAA\cdots$ with two distinct atoms $A$ and $B$, and two independent spring constants $K_{AB}=K_{BA}$ and $K_{AA}$. ($K_{ij}$ is a spring constant between $i$-type atom and $j$-type atom.)
As shown in Fig.\ref{fig:transmittanceassign}, there are four kinds of length 2 supertiles, $ABA,AAB,BAA$ and $BAB$ with the following rules; $BAA$ is followed by $AAB$, whereas $ABA$ is followed by either $BAA$ or $BAB$. Thus, one can express transmittance in terms of $R_U\equiv M_{AAB}M_{BAA}M_{ABA}$ and $R_V\equiv M_{BAB}M_{ABA}$ which correspond to renormalized transfer matrices of $U\equiv ABAAB$ and $V\equiv ABAB$ respectively.
Then, the commutator of $R_U$ and $R_V$ is represented as\cite{macia2000thermal,suck2013quasicrystals},
\begin{align}
\label{eq:resonance}
&[R_U,R_V]=\frac{\lambda}{\gamma}(2\gamma-1-\alpha[1+\lambda(\gamma-1)])\begin{pmatrix} 1 & 0 \\ 2-\alpha\lambda & -1 \end{pmatrix}.
\end{align}
Here, $\alpha=\frac{m_B}{m_A}$, $\gamma$=$\frac{K_{AA}}{K_{AB}}$, and $\lambda=\frac{m_A\omega^2}{K_{AB}}$. Based on Eq.\eqref{eq:resonance}, the critical mode where $R_U$ and $R_V$ commute,  is given for $\lambda = \lambda^*\equiv \frac{\alpha-2\gamma+1}{\alpha(1-\gamma)}$\cite{suck2013quasicrystals}. We first consider the special case $\lambda^*=2$ when $m_A=m_B$ i.e. $\alpha=1$. 
When site $n$ is occupied by $B$, the global transfer matrix is given by, 
\begin{align}
\label{eq:GT}
&\begin{pmatrix} u_{n}  \\ u_{n-1} \end{pmatrix}=(-1)^{n_V}\begin{pmatrix} -1+\gamma^{-1} & 2-\gamma^{-1} \\ -\gamma^{-1} & -1+\gamma^{-1} \end{pmatrix}^{n_U} \begin{pmatrix} u_1 \\ u_0 \end{pmatrix}.
\end{align}
Here, $n_U$ and $n_V$ are numbers of $R_U$ and $R_V$ up to $n$-th site respectively.
Given the Dirichlet boundary condition on the left end (i.e. $u_1=-u_0$), Eq.\eqref{eq:GT} induces the transmittance as,
\begin{widetext}
\begin{eqnarray}
\label{eq:T}
T_n(\cdots B)=\begin{cases} \sech^2\left(\ln|\cos{(n_U\theta)}-\sqrt{2\gamma-1}\sin{(n_U\theta)|}\right) & \mbox{if }\gamma\ge 1/2\\ \sech^2\left(\ln|\cosh{(n_U\phi)}+\sqrt{1-2\gamma}\sinh{(n_U\phi)|}\right) & \mbox{if }\gamma<1/2.
\end{cases}
\end{eqnarray}
\end{widetext}
Here, $T_n(\cdots B)$ defines thermal transmittance when site $n$ is occupied by $B$ (See Fig.\ref{fig:transmittanceassign}.), $\cos{(\theta)}\!=\!\gamma^{-1}\!-\!1$ where $0\!\le\!\theta\!\le\!\pi$ and $\sinh{(\phi)}\! =\!\sqrt{-2\gamma^{-1}\!+\! \gamma^{-2}}$.
When site $n$ is occupied by $A$, the transmittance depends if site $n\!-\!1$  is occupied by either $A$ or $B$. When it is occupied by $B$, the thermal transmittance $T_n(\cdots BA)$ is in similar form with $T_n(\cdots B)$ but replacing  $ \sqrt{\pm (1-2\gamma)}$ into $1/\sqrt{\pm (1-2\gamma)}$. 
However, when it is occupied by $A$  (See Fig.\ref{fig:transmittanceassign}.), thermal transmittance is given as, 
\begin{widetext}
\begin{eqnarray}
\label{eq:T34}
T_n(\cdots AA)=\begin{cases} \sech^2\left(\ln\left|(2\gamma^{-1}-1)\cos(n_U\theta)-\left(\frac{\gamma^{-1}-1}{\sqrt{2\gamma-1}}+\gamma^{-1}\sqrt{2\gamma-1}\right)\sin(n_U\theta)\right| \right), & \mbox{if }\gamma\ge 1/2 \\ \sech^2\left(\ln\left|(2\gamma^{-1}-1)\cosh(n_U\phi)+\left(\frac{\gamma^{-1}-1}{\sqrt{1-2\gamma}}+\gamma^{-1}\sqrt{1-2\gamma}\right)\sinh(n_U\phi)\right| \right), & \mbox{if }\gamma<1/2. \end{cases}
\end{eqnarray}
\end{widetext}
For all cases, there exists a critical strength of quasi-periodicity $\gamma^*=1/2$ which separates two different regimes; (i) critical i.e. neither delocalized nor localized for $\gamma\ge 1/2$ (ii) rapidly decaying i.e. localized for $\gamma<1/2$.
Figs.\ref{fig:inducedFIB} (a) and (b) show two different transmittance characters when $\alpha=1$ for $\gamma=0.501$ and $\gamma=0.499$ respectively.
For $\gamma\ge \gamma^*$, the critical behavior, neither perfectly delocalized nor localized, emerges with a possible fractal structure in the  transmittance. While for $\gamma<\gamma^*$, thermal transmittance decays very rapidly as a function of system size i.e. localized.

 One can further generalize for $\alpha\neq 1$ i.e. $m_A \neq m_B$ and search for the parameter regime showing critical behavior in transmittance. The results are summarized in Eq.\eqref{eq:criticalgamma} where $(x)_{\ge} \left((x)_{\le}\right)$ indicates that the critical state exists when $\gamma\ge x \left(\gamma\le x\right)$ respectively.
Remind that when $\alpha=1=\gamma$, it describes periodic system and perfectly delocalized state is present.
For any $\alpha$, the critical value $\gamma^*=\frac{1}{2}$ exists which separates two regime either critical or localized. 
However, unlike the case $\alpha=1$, there is another finite regime where the localized mode is stabilized for $\gamma$ in between  $\frac{\alpha+1}{2\alpha}$ and $\frac{\alpha+1}{2}$. As $\alpha$ deviates from 1, different masses between $A$ and $B$ make the system harder to transport thermal vibration and this results in additional localized mode at intermediate $\gamma$. 
In terms of the critical state, there are two distinct regimes; 
One is for large $\gamma$ in strong quasi-periodic limit and another is for  intermediate quasi-periodicity with $\frac{1}{2}\le\gamma<1$.  
\begin{align}
\label{eq:criticalgamma}
&\gamma^{*}=\begin{cases} \left(\frac{1}{2}\right)_{\ge} \mbox{ and } \left(\frac{\alpha +1}{2\alpha}\right)_{<} \mbox{ or } \left(\frac{\alpha +1}{2}\right)_{>} & \mbox{if } \alpha>1 \\ \left(\frac{1}{2}\right)_{\ge} \mbox{ and } \left(\frac{\alpha +1}{2}\right)_{<} \mbox{ or } \left(\frac{\alpha +1}{2\alpha}\right)_{>} & \mbox{if } 0<\alpha<1 \\ \left(\frac{1}{2}\right)_{\ge} \mbox{ and } 1_{>,<} & \mbox{if } \alpha=1  \end{cases}
\end{align}

Having explored bizarre properties of critical phonon modes, we now discuss that such character can be classified by PE cohomology of supertilings. To see this explicitly, let's go back to Eqs.\eqref{eq:T} and \eqref{eq:T34}.
The Fibonacci substitution leads renormalized transfer matrices as $R_V\to M_{ABA}R_U$ and $R_U\to M_{ABA}R_VR_U$, where $M_{ABA}$ does not contribute to the renormalized supertiling pattern.
Then, this is equivalent to the case of original Fibonacci tiling and thus its PE cohomology group is simply $\mathbb{Z}^2$ 
\cite{barge2008cohomology,sadun2008topology}.
Explicitly, $n_U$ and $n_V$, numbers of renormalized transfer matrices $R_U$ and $R_V$ respectively, are independent generators of PE cohomology group $\mathbb{Z}^2$.
For $\alpha=1$, $R_V$ is simply an identity matrix and thus transmittance is independent of $n_V$ but only depends on $n_U$. Therefore, the system is classified by the generators $(U,V)= (1,0)\in\mathbb{Z}^2$ and there is only a single finite regime of $\gamma$ where localized mode is stabilized. As $\alpha$ deviates from 1, however, both $n_U$ and $n_V$ contribute in  transmittance. This leads the system to be classified by $(U,V) = (1,0)$ and $(0,1) \in \mathbb{Z}^2$ and generates two $\gamma$ regimes for localized modes.
Hence, the localization character of phonon modes with respect to the quasi-periodic strength is indeed topologically protected and is robust under any kinds of perturbation with PE transformations\cite{kellendonk2015mathematics,sadun2008topology}.

Our approach based on PE cohomology is generally applicable for any quasi-periodic tilings. We can show that the presence of nontrivial critical phonon mode is not guaranteed by PE cohomology. (In supplementary material, exemplifying other types of decorated metallic-mean quasicrystals, we argue that the presence or absence of critical phonon modes depends on the system and is not related to the PE cohomology group.) However, if they exist, the localization behavior of critical modes is indeed classified by the PE cohomology of renormalized supertiling. 
To support our argument, we now explore the Cantor tiling case and show that distinct PE cohomology group results in different localization behavior of thermal transmittance at critical mode frequency. The Cantor tiling is a bi-prototile quasi-periodic system whose quasi-periodicity lies on links; $L,S$. It is generated by the substitutions $L\to LSL$ and $S\to SSS$. Apparently, there are three kinds of supertiles, $X\equiv LS$, $Y\equiv SL$ and $Z\equiv SS$ but no $LL$\cite{jeon2020topological}. In addition, there are certain rules; $Y$ is followed by $X$ and $Z$ always consecutively appears even number of times. Based on the substitution, the first couple of Cantor supertilings are represented as $XYXZ^2YXYXZ^8Y\cdots$ where $Z^k$ is the compact notation for $k$-number of $Z$ supertiles. Then, ignoring the first supertile $X$, the system can be renormalized with two supertiles $YX$ and $Z^2$. The commutator between the renormalized transfer matrices $M_{X}M_Y$ for $YX$ and $M_{Z}^2$ for $Z^2$ is then, 
\begin{align}
\label{Cantorcommutator}
&[M_XM_Y,M_Z^2]=(\lambda-2)\lambda(\gamma^{-1}-1)\sigma_x.
\end{align}
Here, $\lambda=\frac{m\omega^2}{K_S}$ and $\gamma=\frac{K_L}{K_S}$. 
Thus, one can easily read off that the renormalized transfer matrices perfectly commute with each other and the critical mode emerges at $\lambda=2$ no matter what $\gamma$ is. The transmittance where site $n$ is ended with either $X$ or $Y$ or $Z$, is summarizd as following. 
\begin{widetext}
\begin{eqnarray}
\label{eq:transmittanceCantor}
&T_{n}=\begin{cases} \sech^{2}\left(\ln{\left|-3-2N_{YX}+2\gamma^{-1}(N_{YX}+1)\right|}\right), & \mbox{if end with } X \mbox{ or } Z^2 \\ \sech^{2}\left(\ln{\left|-3-2N_{YX}+2\gamma^{-1}(N_{YX}+2)\right|}\right), & \mbox{if end with } Y \\ \sech^{2}\left(\ln{\left|-1+2N_{YX}+2\gamma^{-1}(1-N_{YX})\right|}\right), & \mbox{if end with odd number of } Z.
\end{cases}.
\end{eqnarray}
\end{widetext}
Here, $N_{Z^2}$ and $N_{YX}$ are number of $Z^2$ and $YX$ renormalized supertiles respectively which appear up to $n$-th site. 
In terms of two renormalized supertiles $YX$ and $Z^2$, transmittance only depends on the number of renormalized supertile $N_{YX}$ but not for $N_{Z^2}$ because of $M_Z^2= -I $ at critical mode frequency.
 For all three cases, thermal transmittance of the Cantor tiling rapidly decays as a function of site $n$ regardless the quasi-periodic strength $\gamma$. To understand such unique behavior of the thermal transmittance, we first note that the Cantor supertiling has arbitrarily long consecutive $Z$ supertiles in thermodynamic limit whose lengths are $2^k$ where $k\in \mathbb{Z}$. Hence, in thermodynamic limit, the system can be considered as an almost periodic system composed of $S$ prototile but with small portion of $L$ prototile in between which can be regarded as effective disorders. Thus, it induces rapidly decaying feature of thermal transmittance.

Such rapidly decaying behavior is completely distinct from the transmittance of \textit{induced} Fibonacci quasicrystal and one can understand it based on different PE cohomology groups they belong to. 
To be more specific, we compute the PE cohomology group of  renormalized supertiling in the Cantor tiling and compare it with the PE cohomology group $\mathbb{Z}^2$ for \textit{induced} Fibonacci quasicrystal we have studied.
For the Cantor tiling case, it turns out that transmittance  is characterized by $\mathbb{Z}[1/2]\oplus \mathbb{Z}[1/4]$ where $\mathbb{Z}[1/n]$ represents $\left\{\frac{m}{n^k}: m,k\in\mathbb{Z}\right\}$. (See supplementary material for detailed information.)
This is obviously distinct from the case for \textit{induced} Fibonacci quasicrystal, and is responsible for different transmittance behavior.

\textit{\textbf{Summary and conclusion ---}} \ We study anomalous critical phonon modes and their topological properties in one dimensional quasicrystals. By exemplifying several metallic-mean tilings and the Cantor tiling, the construction of renormalized supertiles and the criteria for presence or absence of critical phonon modes have been studied. 
For certain types of quasi-periodic systems such as \textit{induced} Fibonacci quasicrystal, we show that the critical phonon mode is present and either critical or localized behavior of transmittance is drastically controlled as a function of quasi-periodic strength. Whereas, in different quasi-periodic system such as the Cantor tiling, thermal transmittance always rapidly decays at the critical phonon mode. We emphasize that such bizarre localization behavior of thermal transmittance solely relies on distinct aperiodic tiling patterns and can be classified based on PE cohomology group of supertilings. 

Our new approach based on renormalized supertiles and their PE cohomology group, can be generally applicable for studying one dimensional or even higher dimensional quasicrystals and  their critical phonon modes. It gives a way to classify quasi-periodic systems with  associated critical transmittance characteristics. In addition, similar approach can be also extended to study more complex interactions beyond the nearest-neighbors and to find new types of thermal transmittance behavior which we leave as an interesting future work.

\subsection*{Acknowledgement}
This work is supported by National Research Foundation Grant (NRF-2020R1F1A1073870, NRF-2020R1A4A3079707).

\bibliographystyle{plain}
\bibliography{my1}

\clearpage
\newpage 

\appendix
\renewcommand\thefigure{\thesection\arabic{figure}}
\setcounter{figure}{0}
\setcounter{equation}{0}
\renewcommand{\theequation}{S.\arabic{equation}}
\renewcommand{\thefigure}{S.\arabic{figure}}

\textit{\textbf{Proof of absence of critical phonon mode for any original metallic-mean tilings ---}} Here, we show that original metallic-mean quasicrystals do not have any nontrivial critical phonon mode i.e., the transfer matrices of renormalized supertiles do not commute with each other. In general, the original metallic-mean quasicrystals are generated by the substitution maps $L\to L^{k}S$ and $S\to L$ \cite{senechal1996quasicrystals,kellendonk2015mathematics} and quasi-periodicity lies on links. Especially, the case of $k=1$ is the Fibonacci (or golden-mean) quasicrystal. For clarity, we call the quasicrystal generated by above substitution as $k$-th order metallic-mean quasicrystal. Since all metallic-mean quasicrystals do not have $SS$\cite{senechal1996quasicrystals}, their supertiles used in phonon modes with nearest neighbor interaction are only three types which are the same as the case of original Fibonacci quasicrystal, explicitly $LS,SL$ and $LL$. In addition, since $S$ is substituted into a single $L$, both $SL^kS$ and $SL^{k+1}S$ local patches exist in the $n$-th order metallic-mean quasicrystal. Therefore, consecutive $L$ prototiles appear with either even length or odd length. Thus, in order to renormalize supertiling, we need both $M_{SL}M_{LS}$ and $M_{LL}$ which  is consistent with the case of original Fibonacci quasicrystal. Thus, it is clear that every $k$-th order metallic-mean quasicrystals do not have any critical phonon modes. Similarly, when metallic-mean quasi-periodicity lies on sites instead of links, there is no nontrivial critical phonon mode too.

\textit{\textbf{Critical phonon modes of modified metallic-mean quasicrystals ---}} Beyond \textit{induced} Fibonacci quasicrystal, non-trivial critical phonon modes can exist in metallic-mean tilings with modifications  (especially for silver-mean tiling). To investigate them, we define the $k$-th order \textit{modified} metallic-mean quasicrystal as the system where quasi-periodicity is generated on links by the substitution map $L\to L^{k}S$ and $S\to LL$.
Here, we argue two remarkable characteristics based on PE cohomology group studies. First of all, presence of critical phonon mode is not a topological quantity. Secondly, however, if the critical phonon modes exist, their thermal transmittance behavior is robust under aperiodic tiling space and is protected by PE cohomology group. To state our argument, we exemplify \textit{modified} metallic-mean tilings and study drastic thermal transmittance in these tilings with respect to the strength of quasi-periodicity and classify them with their PE cohomology group. Then, we also discuss even though transmittance of different tilings belong to the same PE cohomology group, presence or absence of critical phonon modes depend on cases.

For the 1st order metallic-mean tiling, consecutive $L$ should appear  either $SLS$ or $SLLLS$. Hence, one can renormalize supertiling into $LSL$ and $LLL$ renormalized supertiles. Corresponding renormalized transfer matrices are $M_{SL}M_{LS}$ and $M_{LL}^2$ respectively. Thus, from the commutation relation between $M_{SL}M_{LS}$ and $M_{LL}^2$, we obtain nontrivial critical phonon mode for $\lambda\!=\!\frac{m\omega^2}{K_L}\!=\!2$. For $\lambda\!=\!2$, we have $M_{LL}^2\!=\!-I$ where $I$ is an identity matrix and $M_{SL}M_{LS}=\begin{pmatrix} \gamma^{-1}-2 & -1+\gamma^{-1} \\ 1-\gamma^{-1} & -\gamma^{-1} \end{pmatrix}$. Given Dirichlet boundary condition for the left end ($u_1=-u_0$), perfect transmittance is derived except the case where link between site $n$ and site $n-1$ is $S$ type. Explicitly, thermal transmittance is given by,
\begin{align}
\label{TTT}
&T_n=\begin{cases} \sech^2\left(\ln\gamma^{-1}\right) & \mbox{if } [n-1,n]=S \\ 1 & otherwise. \end{cases}
\end{align}
Unlike \textit{induced} Fibonacci quasicrystal, the 1st order \textit{modified} metallic-mean tiling does not exhibit drastic changes of localization in thermal transmittance and this can be understood based on the PE cohomology of this tiling. 
To discuss it, let's first see how the substitution maps are induced with the renormalized supertiles. From the substitution maps for $L$ and $S$, one can construct the substitution matrix as $\mathcal{S}=\begin{pmatrix} 1 & 2 \\ 1 & 0 \end{pmatrix}$ whose basis is composed of the number of renormalized supertiles $LSL$ and $LLL$ respectively. 
Its PE cohomology group is $\mathbb{Z}[\frac{1}{2}]\oplus\mathbb{Z}$ which can be obtained from the direct limit of $\mathcal{S}^T$\cite{kellendonk2015mathematics,sadun2008topology}. 
Thermal transmittance in Eq.\eqref{TTT} does not depend on any generators of PE cohomology group i.e number of renormalized supertiles $LSL$ and $LLL$, thus the system is classified with a trivial element, $(0,0)\in\mathbb{Z}[\frac{1}{2}]\oplus\mathbb{Z}$. 
It is indeed distinct from $\mathbb{Z}^2$ which is the case for  \textit{induced} Fibonacci quasicrystal. We emphasize that entire PE cohomology groups are different between these two cases and this causes totally distinct thermal transmittance behaviors with respect to the strength of quasi-periodicity, $\gamma$.

Now, let's consider the 2nd order \textit{modified} metallic-mean quasicrystal. In this case, $L$ appears successively, and hence there is no $SLS$ local patch. In addition, the substitution $L\to LLS$ and $S\to LL$ is very similar to the substitution rule of the original Fibonacci quasicrystal, thus similar forbidden rules of local patches can be applied having no $LLLLLL$ and etc\cite{senechal1996quasicrystals}. In addition, $L$ prototile appear only in the form of either $SLLS$ or $SLLLLS$. Therefore, one can represent the tiling into two renormalized supertiles; $LLSL$ and $LLL$. Then, the commutation relation between the transfer matrices of these two supertiles are written as following.
\begin{align}
\label{commutatormodified}
&[M_{SL}M_{LS}M_{LL},M_{LL}^2]=\lambda(1-\gamma^{-1})(2-\lambda)\begin{pmatrix} 1 & 0 \\ 2-\lambda & -1 \end{pmatrix},
\end{align}
where $\gamma=\frac{K_S}{K_L}$ and $\lambda=\frac{m\omega^2}{K_L}$. From Eq.\eqref{commutatormodified}, one can read off the presence of  critical phonon mode when $\lambda=2$. Given Dirichlet boundary condition for the left end ($u_1=-u_0$), thermal transmittance in this case shows exactly same as $\alpha=1$ case of \textit{induced} Fibonacci quasicrystal. One can further check that $M_{SL}M_{LS}M_{LL}$ and $M_{LL}^2$ equivalent to $R_U$ and $R_V$ renormalized transfer matrices defined in the case of \textit{induced} Fibonacci quasicrystal with $\alpha=1$ in a sense that $S=AA$ and $BA=L=AB$. Hence the two systems are indeed equivalent to each other.
Specifically, critical quasi-periodic strength that separates either critical or localized behavior in thermal transmittance also appears in the 2nd order \textit{modified} metallic-mean tiling.
This can be understood by PE cohomology equivalence between the 2nd order \textit{modified} metallic-mean tiling and $\alpha=1$ case of \textit{induced} Fibonacci quasicrystal. 
We note that the two tilings have exactly equivalent local matching rules of supertiling, thus share the same tiling space topology and results in the same thermal transmittance behavior at the critical mode\cite{sadun2008topology, kellendonk2015mathematics,senechal1996quasicrystals}.

We now consider general $k$-th order \textit{modified} metallic-mean quasicrystals with $k>2$. First note that in this case, consecutive $L$ prototiles appear with length either $k$ or $k+2$. Thus, consecutive $M_{LL}$ transfer matrices appear with length of either $k-1$ or $k+1$ and the global transfer matrix is split into $M_{SL}M_{LS}M_{LL}$, $M_{LL}^{k-2}$ and $M_{LL}^{k}$. This leads the transfer matrix of supertiling to be  renormalized by $M_{SL}M_{LS}M_{LL}$ and $M_{LL}^{gcd(k,k-2)}$ where $gcd(x,y)$ is greatest common divisor of $x$ and $y$. For the case of even number $k$, $gcd(k,k-2)=2$ while for the case of odd number $k$, $gcd(k,k-2)=1$. Thus, we can divide into two cases; (i) $k$ is even (ii) $k$ is odd. When $k$ is even number, commutator of renormalized transfer matrices is exactly same as one of the 2nd order \textit{modified} metallic-mean tiling. Thus, the resultant transmittance is also equivalent to the case of the 2nd order \textit{modified} metallic-mean tiling. Explicitly, they have nontrivial critical phonon mode at $\lambda=\frac{m\omega^2}{K_L}=2$ and the transmittance shows the critical behavior for $\gamma\ge0.5$, $\gamma\neq1$ regime as we discussed. While, for odd $k$ with $k>2$, the commutator between the renormalized transfer matrices is exactly same as the case of original metallic-mean quasicrystals. Hence, for odd $k$ with $k>2$, the $k$-th order \textit{modified} metallic-mean tiling do \textit{not} have any critical phonon mode.
However, for all $k>2$, PE cohomology groups of the $k$-th order \textit{modified} metallic-mean supertilings are identical to be $\mathbb{Z}^2$. To be more specific, the substitution matrix for each $k$ is given as 
$\mathcal{S}_{k, even}=\begin{pmatrix} k-1 & k-1 \\ 1 & 1 \end{pmatrix}$ and $\mathcal{S}_{k, odd}=\begin{pmatrix} 2k-2 & 2k \\ 1 & 1 \end{pmatrix}$ respectively. (Here the basis is chosen to be the numbers of renormalized supertiles $LLSL$ and $L^{gcd(k,k-2)+1}$ respectively.) Thus, the direct limit of transpose of these matrices is classified by $\mathbb{Z}^2$ and this results in PE cohomology group of supertiling for $k$-th order \textit{modified} metallic-mean quasicrystals is $\mathbb{Z}^2$.
Based on above argument, we emphasize that presence of nontrivial critical phonon mode is \textit{not} topologically guaranteed. In other words, although different tilings share the same PE cohomology group, it is possible that only parts of them have nontrivial critical phonon modes.

\textit{\textbf{PE cohomology group of the Cantor supertiling ---}} \ Here, we explicitly derive PE cohomology group of the renormalized Cantor supertiling. This work is derived using the method given in Refs \onlinecite{barge2008cohomology,sadun2008topology,kellendonk2015mathematics,jeon2020topological,varjas2019topological}. For clarity, here, we use the notation $N_*$ for the (integrated) 1-cochain map on the supertiling which counts the number of $*$-type (renormalized) supertile.

\begin{figure}[!h]
\centering
\includegraphics[width=0.5 \textwidth]{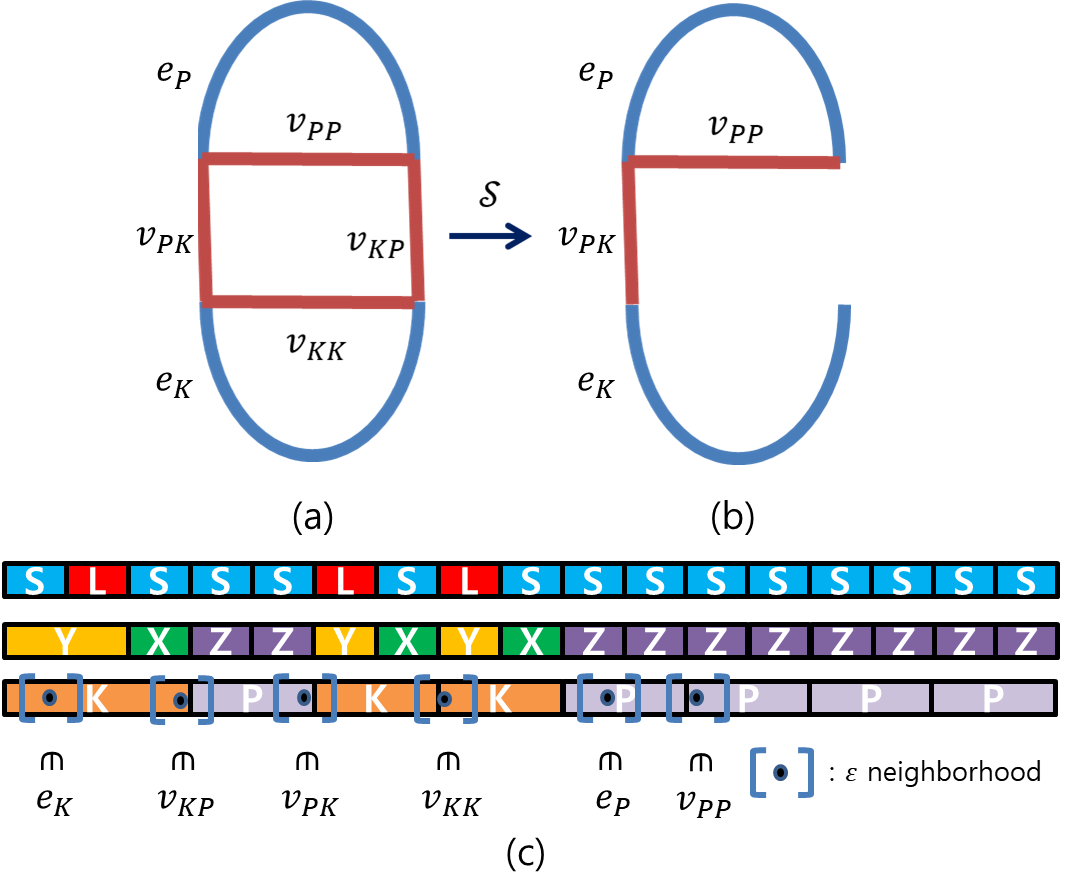}
\caption{\label{fig:BD} Barge-Diamond (BD) complex of the renormalized Cantor supertiling under the substitution map given by Eq.\eqref{eq:subcantor}; (a) BD complex for the renormalized Cantor supertiling and (b) BD complex in the eventual range of vertex flips.  $e_{P}$ and $e_{K}$ represent equivalence classes of positions on the tiling whose neighborhood (within the small size $\varepsilon$, for instance) is completely confined within the same renormalized supertiles $P$ and $K$ respectively. $v_{ij}, i,j=P \ \mbox{or} \ K$, called a vertex flip, represents equivalence classes of positions whose $\varepsilon$ size neighborhood contains intersection of $i$ and $j$ renormalized supertiles. (c) Illustration of such equivalence classes of positions defined in the renormalized Cantor supertiling with $\varepsilon$-neighborhood at each position. Remind that supertiles $X,Y$ and $Z$ are taken to overlap each prototile. For instance, $Y$-$X$-$Z$ $\cdots$ at the beginning of array is for $SL$-$LS$-$SS$ where each prototile is overlapped for both left and right sides of the supertiles. In order to get the tiling cohomology in thermodynamic limit, it is known that the BD complex in the eventual range under substitution map should be considered  \cite{sadun2008topology,jeon2020topological}.}
\end{figure}

First of all, based on the substitution maps of the Cantor tiling, the substitution maps of renormalized supertiles $K\equiv YX=SLS$ and $P\equiv ZZ=SSS$ are given by $K\to PKKP$ and $P\to PPPP$ respectively. Thus, one can construct the substitution matrix, $\mathcal{S}$,
\begin{align}
\label{eq:subcantor}
&\mathcal{S}=\begin{pmatrix} 2 \ 0  \\ 2 \ 4 \end{pmatrix}.
\end{align}
Here, the ordered basis is chosen to be the numbers of $K$ and $P$ renormalized supertiles respectively. 
Then, to classify cochain maps of the supertiling, we construct the Barge-Diamond (BD) complex, the complex composed of equivalence classes of position on the tiling based on their local patterns\cite{barge2008cohomology,sadun2008topology, jeon2020topological}. In our case, there are four kinds of vertex flips, $v_{ij}$, which are the equivalence classes of positions whose small neighborhood with a finite size contains intersection of both $i$ and $j$  renormalized supertiles\cite{sadun2008topology,jeon2020topological}; $v_{KK},v_{KP},v_{PK},v_{PP}$ as shown in Fig.\ref{fig:BD} (c). By applying the substitutions of the original Cantor tiling, i.e. $L\to LSL, S\to SSS$ to the vertex flips, one can show that only $v_{PK}, v_{PP}$ are in the eventual range\cite{kellendonk2015mathematics,sadun2008topology} of the substitution maps. Explicitly, under the substitution maps of $L,S$ in the original Cantor tiling, $v_{KP},v_{PP}\to v_{PP}$ and $v_{KK},v_{PK}\to v_{PK}$. Hence, these two among four kinds of vertex flips contribute to the tiling space topology in thermodynamic limit\cite{sadun2008topology}. There are other types of equivalence classes of positions whose small neighborhood with a finite size does not contain any intersection of different supertiles but is completely confined into a renormalized supertiles $P$ or $K$. We denote such equivalence classes by $e_P$ and $e_K$ respectively. (See Fig.\ref{fig:BD} (c))  Figs.\ref{fig:BD} (a) and (b) illustrates the BD complex and its eventual range of the substitution map respectively.

\begin{figure}[!h]
\centering
\subfloat[]{\includegraphics[width=0.5 \textwidth]{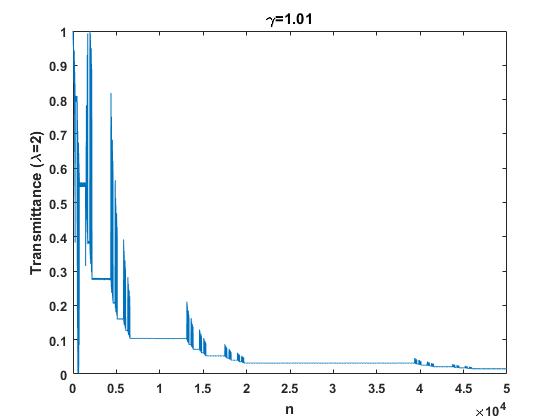}
\label{fig:1.01Cantor}}
\hfill
\subfloat[]{\includegraphics[width=0.5 \textwidth]{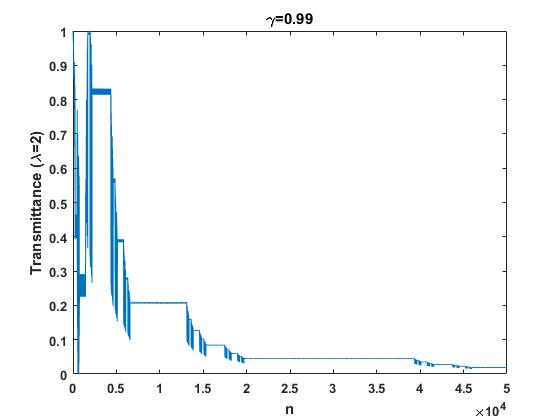}
\label{fig:0.99Cantor}}
\caption{\label{fig:Cantor} Thermal transmittance of the Cantor tiling for $\gamma = 1.01$ (a) and $\gamma=0.99$ (b) at $\lambda=\frac{m\omega^2}{K_S}=2$. Both cases exhibit decaying behavior i.e., localized phonon mode appears. The regions where transmittance is constantly maintained,  transmittance appear due to the presence of periodic (sub)pattern by $S$ prototiles. In addition, self-similar pattern also appears regardless of specific $\gamma$ values. Sharp peaks or drops are originated from $K=YX$ renormalized supertile (or equivalently presence of $L$ prototiles which can be understood as effective disorders). They appear in self-similar way and the feature of sharp peak or drop depends on the sign of $\gamma-1$ changing the relative ratio between $K_L$ and $K_S$. 
}
\end{figure}

Considering the BD complex in the eventual range (Fig.\ref{fig:BD} (b)) and the direct limit\cite{kellendonk2015mathematics,sadun2008topology} of $\mathcal{S}^T$ where $\mathcal{S}$ in Eq.\eqref{eq:subcantor}, one can conclude that the PE cohomology group of the renormalized Cantor supertiling is $\mathbb{Z}[\frac{1}{2}]\oplus \mathbb{Z}[\frac{1}{4}]$.
To be more specific, let's consider left eigenvectors of $\mathcal{S}^T$,  $v_2=\begin{pmatrix} 1 & -1 \end{pmatrix}$ and $v_4=\begin{pmatrix} 0 & 1 \end{pmatrix}$ which correspond to $N_{K}-N_{P}$ and $N_P$ with the eigenvalues 2 and 4 respectively\cite{kellendonk2015mathematics,jeon2020topological}. 
In other words, $N_{K}-N_{P}$ and $N_P$ are nothing but the generators with element $1\in\mathbb{Z}[\frac{1}{2}]$ and $1\in\mathbb{Z}[\frac{1}{4}]$ respectively. Thus, $N_{K}$ which is a linear combination of them belongs to $\mathbb{Z}[\frac{1}{2}]\oplus \mathbb{Z}[\frac{1}{4}]$ distinct from $\mathbb{Z}^2$ as $(K-P,P)=(1,1)\in\mathbb{Z}[\frac{1}{2}]\oplus \mathbb{Z}[\frac{1}{4}]$. 
Thus, it is obviously different from the PE cohomology group of \textit{decorated} metallic-mean quasicrystals and one can conclude that the critical phonon mode of the Cantor tiling behaves differently. In this case, the critical phonon mode of the Cantor tiling exhibits localized transmittance regardless of quasi-periodic strength (See Fig.\ref{fig:Cantor}).

Figs.\ref{fig:Cantor} (a) and (b) show thermal transmittance of the critical mode for the Cantor tiling with $\lambda=\frac{m\omega^2}{K_S}=2$. No matter what quasi-periodic strength, $\gamma$, is, one can observe transmittance rapidly decays i.e. localized. However, there are certain regions where transmittance is not decaying but constantly maintained. It is  originated from consecutive long range patches composed of $S$ prototiles only, i.e. consecutive $P$ renormalized supertiles. In addition, for any $\gamma$, the thermal transmittances explicitly exhibit self-similar structure which can be understood by the unique self-similar pattern of the Cantor tiling\cite{kellendonk2015mathematics}.

\end{document}